\documentstyle[12pt,psfig,epsfig]{article}
\newcommand{\beq}{\begin{equation}}
\newcommand{\eeq}{\end{equation}}
\newcommand{\beqarray}{\begin{eqnarray}}
\newcommand{\eeqarray}{\end{eqnarray}}
\def\lsim{\raise0.3ex\hbox{$\;<$\kern-0.75em\raise-1.1ex\hbox{$\sim\;$}}}
\def\gsim{\raise0.3ex\hbox{$\;>$\kern-0.75em\raise-1.1ex\hbox{$\sim\;$}}}
\def\para{\vspace{0.3cm}\noindent}

\begin{document}
\begin{center}
{\large \bf A study of the appearance of tau neutrinos from a gamma ray burst  by detecting their horizontal electromagnetic showers}

\medskip

{Nayantara Gupta \footnote{bunkagupta@yahoo.com}}\\  
{\it Department of Theoretical Physics,\\
 Indian Association for the Cultivation of Science,\\
Jadavpur, Kolkata 700 032, INDIA.}  
\end{center}

\begin{abstract}

We explore the possibilty of detecting horizontal electromagnetic showers of
 tau neutrinos from individual gamma ray bursts, in large scale detectors like
 HiRes and Telescope Array. We study the role of the parameters of a gamma ray
 burst in determining the expected number of tau events from that burst. The 
horizontal beam of tau leptons produce visible signals in the atmosphere. We
 find that there is a slim chance of observing tau lepton
 appearances from GRBs with Telescope Array. The number of signals is strongly
 dependent on the Lorentz factor $\Gamma$, redshift $z$ of a GRB, energy 
emitted in muon neutrinos and antineutrinos $E_{\nu,HE}$ and also on some
 other parameters of a GRB. It is possible to understand neutrino oscillations
 in astrophysical neutrinos and the mechanism or model of neutrino production
 inside a GRB by detection or non detection of tau lepton signals from it.
  
\end{abstract} 

\newpage
\section{Introduction}
The very high energy muon and electron neutrinos produced inside an astrophysical 
source cannot pass through the matter of Earth due to their interaction with
 the matter of Earth and produce signals in a detector on Earth's surface. The
 very high energy upward going muon and electron neutrinos are shadowed by
 Earth. However, very high energy tau neutrinos can penetrate the Earth to some extent. In models of neutrino production from proton acceleration inside an astrophysical source the intrinsically produced tau neutrino flux is expected to be very small. But tau neutrinos can be generated by vacuum flavour oscillations of muon neutrinos.  
There is a possibility of investigating the  appearance of tau neutrinos from 
an astrophysical source of very high energy muon neutrinos by detecting these 
tau neutrinos.
The expected number of events from upward going high energy tau neutrinos of a
 gamma ray burst in a muon detector has been calculated in \cite{nayan1}.

\para

The suggestion of detecting tau neutrino interaction through the shower 
induced by  tau lepton decay in the atmosphere is given in \cite{fargion,selvon}
.
If the neutrinos are passing through the surface of Earth almost horizontally
 the charged leptons produced by them are also emitted in the horizontal 
direction. The electrons produced from a very high energy horizontal beam of 
electron neutrinos do not escape the matter of Earth. The muons produced from
 very high energy muon neutrinos by their interaction with the matter of Earth
 do not produce visible signals in the atmosphere as the electromagnetic halo 
that surrounds very high energy muons does not spread enough in space to
 produce detectable signals in arrays of detectors separated by distances of
 the order of a kilometer.
 The tau leptons produced from very high energy tau neutrinos are capable of
 producing clear signals if they decay above the detector \cite{bertou,feng}.
 However, the detectability of the visible signals from high energy tau 
neutrinos depends on the detector geometry \cite{bertou}. 
We need an almost horizontal beam of high energy tau neutrinos from the source
 and the neutrino energies as well as the distances between the point of 
interactions and the detector must match to satisfy the condition of observing
 tau decays with ground level fluorescence detectors.  
In \cite{bertou} they have shown that $90\%$ of the detectable signal in
 Auger detector comes from upward going tau neutrinos. In this case the 
interactions occur in the ground all around the array. The downward going tau
 neutrinos interact in the mountains surrounding the array and generate $10\%$
 of the detectable signal.
In \cite{sasaki} they have discussed how Telescope Array can explore very
 high energy cosmic neutrinos by detecting the earth-skimming tau leptons using
 a large array of bright and wide field-of-view fluorescence telescopes.

\para

At present there are several models which have been proposed to explain the
 occurence of a GRB. There are suggestions that a GRB may evolve from a merger
 of a binary system of compact objects. This binary system may include two 
neutron stars \cite{eichler}, a neutron star and a black hole \cite{narayan}, a
 black hole and a Helium star or a white dwarf \cite{fryer1,fryer2}.
A short duration GRB may originate from the merging of two neutron stars or 
from the merger of a black hole and a neutron star. 
Collapsars, helium mergers and white dwarf-black hole binary mergers may 
give birth to long duration GRBs. When a black hole is created and accretion at
 a rapid rate from the surrounding accretion disk feeds a strong relativistic jet
 in the polar regions, a GRB can arise \cite{woosley, pacy}. In \cite{vietri} 
the authors have suggested a scenario in which a supramassive neutron star 
loses so much angular momentum that centrifugal support against self-gravity
 becomes impossible, and as a result the star implodes to a black hole.  
The supernova explosion in which a supramassive neutron star is formed
sweeps the medium surrounding the remnant, and the quickly spinning remnant
loses energy through magnetic dipole radiation at a rate exceeding its Eddington
luminosity by four orders of magnitude. Occasionally this model may produce 
quickly  decaying or non-detectable afterglows.
This is called the supranova model for gamma-ray bursts. All of the GRB models
 discussed here produce fireballs. The crucial point of the fireball model is that it channels 
a large explosion energy $(>10^{53}erg)$ with small baryon contamination
 \cite{rees}. In the environment provided by the supranova model 
of GRBs \cite{vietri} the baryon contamination is less compared to that provided by the
 collapsar/hypernova model of GRBs.  

\para

The observed spectrum of GRBs and their afterglows are well described by the
 synchrotron and inverse Compton emission of relativistic electrons \cite{granot}.
 It is better studied during the afterglow stage, where we have broad band
 observations. Within the fireball model, both the GRBs and its afterglows are
 due to the deceleration of a relativistic flow.
If there are also relativistic protons in the fireball their interaction with 
the low energy photons may generate high energy neutrinos. Protons can be shock 
accelerated to energies $10^{21}$ eV by the Fermi mechanism and they may 
interact with the photons of the fireball to produce pions which will 
subsquently decay to produce neutrinos \cite{waxman}.
In this work we restrict ourselves only to the fireball model and production 
of neutrinos from photo-pions.

The association of GRBs with the production of ultra high enrgy cosmic rays
 above the GZK cutoff can not be ruled out yet \cite{vietri1}. In the future
 HiRes \cite{hires}, AGASA ({\it Akeno\ Giant\ Air\ Shower\ Array})
 \cite{agasa}, Auger Observatory \cite{auger}, Telescope Array \cite{sasaki}
 etc. would detect more ultra high energy cosmic ray events and help us to
 reveal the unknown facts behind their origin and propagation. 

\para

For a point source like an individual gamma ray burst the nadir angle at which
 it will be observed from Earth is fixed at a particular instant of time.
 Gamma ray bursts are generally of a few seconds durations and therefore the
 position of a burst with respect to an observer on Earth will not change 
significantly during the burst. We can expect to detect horizontal 
electromagnetic showers on Earth from tau neutrinos of a gamma ray burst if
 the burst takes place near the horizon.

The detection of visible signals from horizontal showers of tau neutrinos from
 a GRB with a ground array of detectors is not possible unless the burst
 happens on the night sky and the tau neutrinos are observed within $5^{\circ}$ 
from the horizon. A tau lepton emerging at an angle more than 0.3 rad above the 
horizon would not produce an observable shower at ground level \cite{bertou}.
 
\para

We carry out a study on the detectability of horizontal tau neutrino showers 
from individual gamma ray bursts in the present work. The dependence of the
 number of visible tau neutrino events on the GRB parameters has been explored. The advantage of doing this study is that one can estimate the expected number of such events for a future burst from the figures of our present work.

\section{The parameters of a GRB in the fireball model and the neutrino spectrum from a GRB}
Observations imply that GRBs are produced by the dissipation of the kinetic energy of a highly relativistic fireball. Protons accelerated to high energies in the fireball lose energy through photo-meson interaction with the fireball photons.
The photo-mesons produced by proton photon interaction can subsequently decay to neutrinos.  We define the luminosity of this photon radiation from a GRB by $L_{\gamma}$. Current models of GRBs like the collapsar or hypernova model are capable of producing total energy outputs of the order of $10^{53} erg$ to $10^{54} erg$. However, only a few percent of this energy may be actually available to accelerate ultra high energy protons and subsequently, about $10\%$ of the accelerated proton energies will be transferred to the muon neutrinos and antineutrinos.
In this section we will from time to time refer to three reference frames, (1)
wind rest frame, which is the rest frame of the relativistic wind moving with
 Lorentz factor $\Gamma$. (2) observer's rest frame, in which ultrarelativistic
 outflow moves with an average Lorentz factor $\Gamma$ and (3) the reference
 frame of the observer on Earth, in which the effects of light travel come
 into play.

 The total energy emitted by a GRB in muon neutrino and antineutrino emission is denoted by $E_{\nu,HE}$ in the observer's rest frame, its redshift by $z$ and its Lorentz factor by $\Gamma$. The photon spectrum break energy in MeV is $E_{\gamma,MeV}^{b}$ and the variability time of the source is $t_v$ seconds in the observer's rest frame , which is without the correction due to redshift of the source.
$\epsilon_{e}$ is the fraction of proton's shock thermal energy transformed to electrons and $\epsilon_{B}$ is the fraction of proton's shock thermal energy carried by magnetic field in the fireball model \cite{guetta1}. 

The maximum cutoff energy of the muon neutrino spectrum is $E_{{\nu}max}$ at the source in the observer's rest frame.
The observed maximum cutoff energy of the muon neutrino spectrum on Earth is
$E_{{\nu}max,ob}=E_{{\nu}max}/(1.+z)$, where $z$ is the redshift of the GRB. 
Protons can be accelerated by the Fermi Mechanism to energies $10^{21} eV$ \cite{waxman1} inside a fireball and about $10\%$ of that energy gets transmitted to the secondary muon neutrinos and antineutrinos. Hence we can assume $E_{{\nu}max}$
to be $10^{11} GeV$.
If the Lorentz factor variability within the wind is significant, internal shocks would reconvert a substantial part of the kinetic energy to internal energy. The internal energy may then be radiated as $\gamma$-rays by synchrotron and inverse Compton emission of shock-accelerated electrons. The internal shocks are expected to be mildly relativistic in the wind rest frame. This is due to the fact that the allowed range of shell Lorentz factors is $10^{2}$ to $10^{3}$, implying that the Lorentz factors associated with the relative velocities are not very large.

The observed photon spectrum on Earth from a GRB can be expressed as a broken power law. The data obtained by BATSE (Burst and Transient Source Experiment) can be fitted using the parametrisation given by eqn.(1).

\beq
\frac{d n_{\gamma}}{d \epsilon_{\gamma,ob}} \propto \left\{ \begin{array}{r@{\quad \quad}l}
{\epsilon_{\gamma,ob}}^{-\alpha-1} & \epsilon_{\gamma,ob}<\epsilon_{\gamma,ob}^{b}\\ {\epsilon_{\gamma,ob}}^{-\beta-1} & \epsilon_{\gamma,ob}>\epsilon_{\gamma,ob}^{b}
\end{array} \right.  
\eeq

$\epsilon_{\gamma,ob}$ is the photon energy and $\epsilon_{\gamma,ob}^{b}$ is the break energy of the photon spectrum as measured on Earth.
In analogy with the observed photon spectrum from a GRB the neutrino spectrum from that GRB is expressed below including the effect of energy loss via synchrotron emission by the highest energy pions \cite{guetta}. The neutrino spectrum in the observer's rest frame is given below.

\beq
E_{\nu}^2 \frac{dN_{\nu}}{dE_{\nu}}\propto \left \{ \begin{array}
{l@{\quad \quad}l}
(E_{\nu}/E_{\nu}^{b})^{\beta}&E_{\nu}<E_{\nu}^{b} \\ (E_{\nu}/E_{\nu}^{b})^{\alpha}&E_{\nu}^{b}<E_{\nu}<E_{\nu}^{s} \\ (E_{\nu}/E_{\nu}^{b})^{\alpha}(E_{\nu}/E_{\nu}^{s})^{-2}&E_{\nu}>E_{\nu}^{s}
\end{array} \right.   
\eeq

In the observer's rest frame there is a threshold energy for pion production in proton-photon interaction. 
For the energy of the proton-photon interaction to exceed the threshold energy for the $\Delta$ resonance, the proton energy in the observer's rest frame must
satisfy the condition,
 
\beq
E_{p}^{b}\ge (1.4/{\epsilon^{b}_{\gamma, MeV}}){\Gamma}^{2}_{2.5}\times 10^{7} GeV
\eeq
 
From the expression of proton break energy one can obtain the neutrino spectrum break energy. We assume that on the average $20\%$ of the initial proton energy is transformed to the produced pion. We also assume that the four final state leptons produced in the decay chain $\pi^{+}\rightarrow \nu_{\mu} {\mu}^{+}\rightarrow \nu_{\mu} e^{+} \nu_{e} \bar \nu_{\mu}$ equally share the pion energy.
Then the  expression for neutrino spectrum break energy in the observer's rest frame is as follows

 \beq
E_{\nu}^{b} =7\times 10^{5}\frac{\Gamma_{2.5}^{2}}{\epsilon_{\gamma,MeV}^{b}} GeV.
\eeq 

where, $\Gamma_{2.5}=\Gamma/10^{2.5}$. 
It is assumed that the wind luminosity carried by internal plasma energy $L_{int
}$, is related to the observed $\gamma$-ray luminosity through 
$L_{int}=L_{\gamma}/{\epsilon_{e}}$. This assumption is justified because the electron synchrotron cooling time is short compared to the wind expansion time and hence electrons loose all their energy radiatively.
 
The expression for photon spectral break energy is 
\beq
\epsilon_{{\gamma},MeV}^{b}\approx {\epsilon_{B}}^{1/2}{\epsilon_{e}}^{3/2}\frac{L_{{\gamma},52}^{1/2}}{{\Gamma}_{2.5}^{2}t_{v,-2}}
\eeq

Details of the steps leading to the above expression can be looked up in
 \cite{waxman1}.
Although at present there is no theory that allows the determination of the
 values of the equipartition fractions $\epsilon_{e}$ and $\epsilon_{B}$, one
 can see  from eqn.(5) that fractions not far below unity are required to
 explain the observed spectrum of $\gamma$-rays from GRBs.
The highest energy pions may lose some energy via synchrotron emission before
 decaying, thus reducing the energy of the decay neutrinos. This effect
 becomes important when the pion lifetime becomes comparable to the sychrotron
 loss time.One can compare the synchrotron loss time with the time over which
 pions decay.

In eqn.(2) $E_{\nu}^{s}$ is the neutrino energy above which the synchrotron energy loss of the high energy pions which have decayed to generate the neutrinos becomes important.
Assuming that a neutrino or an antineutrino takes only one fourth of the pion energy one can write down the expression for $E_{\nu}^{s}$. 

\beq
E_{\nu}^{s}=10^{8}\epsilon_{e}^{1/2}\epsilon_{B}^{-1/2}L_{\gamma,52}^{-1/2}\Gamma^{4}_{2.5}t_{v,-2} GeV
\eeq

A detailed derivation of the above expression could be seen in \cite{guetta}.
 The spectrum of muon neutrino and antineutrino from a GRB can be normalised
 using the information of the total energy emitted by that GRB in muon neutrino
 and antineutrino emissions ($E_{\nu,HE}$).

\beq
E_{\nu,HE}=\int_{E_{{\nu}min}}^{E_{{\nu}max}} E_{\nu} \frac{dN_{\nu}}{dE_{\nu}} dE_{\nu}
\eeq
In our case the minimum neutrino energy $E_{{\nu}min}$ is much less than the
 neutrino spectrum break energy $E_{\nu}^{b}$.
 We assume that a GRB at a redshift of $z$ is emitting neutrinos isotropically
 to obtain the expected number of neutrinos on Earth from that GRB.
The observed number of muon neutrinos and antineutrinos on Earth
 per unit area of the surface of Earth and unit energy at observed energy
 $E_{{\nu}obs}$ is
\beq
\frac{d{M}_{{\nu}obs}}{dE_{{\nu}obs}}=\frac{dN_{\nu}}{dE_{\nu}}\frac{1}{4\pi d^2(z)} (1+z).
\eeq

$d(z)$ is the comoving radial coordinate distance of the source. In a spatially
 flat universe we calculate the comoving distance of a source using 
$\Omega_{\Lambda}=0.73$, $\Omega_{m}=0.27$ and $H_o=71 km sec^{-1} Mpc^{-1}$
 from \cite{wmap}.

\para
If the fireball is not spherically symmetric and emits particles in a cone,
 then that cone may face towards or away from the direction of Earth. If the
 emissions from a GRB are not towards Earth we will not detect any signal from
 it. Suppose a jet-like fireball facing towards Earth has a jet opening angle
 $\phi_{jet}$. We denote the source size by $r$. The fireball is expanding
 relativistically. The time required for significant source expansion 
corresponds to a comoving time $t_{co}\sim{r/{c{\Gamma}}}$. One can see 
\cite{waxman1} for discussions on fireball geometry for a jet-like fireball.
The linear size of causally connected regions is $ct_{co}\sim {r/{\Gamma}}$ in
 the reference frame of the fireball, from which it follows that the angular
 size of the causally connected regions is $c t_{co}/r \sim {\Gamma}^{-1}$.
As a result of relativistic beaming an observer can see only a limited portion
 of a GRB with angular size $\sim {\Gamma}^{-1}$. 
If the jet opening angle of a fireball, $\phi_{jet}$ is greater than
 ${\Gamma}^{-1}$ then due to relativistic beaming of emission a distant
 observer cannot distinguish between a sphericall fireball and a jet-like
 fireball. As long as Lorentz factor of the fireball $\Gamma$ remains
 sufficiently
 large such that the wind is ultrarelativistic ($\Gamma \sim 300$) the
 expression obtained for the neutrino spectrum assuming a spherical fireball
 can also be used if the fireball is jet-like.

\para

The relativistic beaming has two effects \cite{frail}. If the beaming factor 
is $b$ then the energy emitted by a GRB is smaller by a factor of $b$ than the
 energy emitted assuming an isotropic emission of energy. The second effect is
 that the actual GRB rate is larger than the observed GRB rate by a factor of
 $b$.

\para
We ultimately want to know the tau neutrino spectrum from a GRB due to
 $\nu_{\mu}\rightarrow \nu_{\tau}$ oscillations.
The probablity of vacuum flavour oscillation from
 $\nu_{\mu}$ to $\nu_{\tau}$ is 

\beq
Prob(\nu_{\mu} \rightarrow \nu_{\tau})=sin^{2}{2 \theta} sin^{2}(\frac{\delta m^2 L}{4 E_{\nu}}).
\eeq
 where $L$ is the distance the neutrinos travel from the source to the detector.
Super-Kamiokande \cite{super} results on $GeV$ energy atmospheric neutrinos
 give us the following best fitted values for mass difference ($\delta m^2$) 
and mixing parameter ($sin^2{2 \theta}$) for $\nu_{\mu} \rightarrow \nu_{\tau}$
 oscillations.

\beq
\delta m^2 \approx 10^{-3}eV^2, sin^2{2 \theta}\approx 1.
\eeq

If the source is at a distance of a megaparsec to thousands of megaparsecs 
away from us the above expression of oscillation probablity averages to about
 half for all relevant neutrino energies to be considered for detection.

If we denote  the number of muon neutrinos by $M_{\nu_{\mu}}$ and the number
 of tau neutrinos produced by oscillation of muon neutrinos by $M_{\nu_{\tau}}$.
Then they are related as described below 

\beq
 M_{\nu_{\tau}}= M_{\nu_{\mu}}\times Prob(\nu_{\mu}\rightarrow \nu_{\tau}).
\eeq

In this way we obtain the expected number of tau neutrinos from a GRB produced
 due to vacuum oscillations of muon neutrinos.
Originally there were almost no tau neutrinos in the GRB compared to the number
 of muon and eletron neutrinos. If we can detect tau neutrinos from a GRB by
 large scale experiments it will be a tau appearance experiment. Since they 
are very high energy tau neutrinos,  they will preserve the directionality of 
the source.

\section{The tau neutrino signals visible near the horizon of Earth}

We follow the procedure discussed in \cite{feng} to calculate the number 
of tau neutrino signals in a ground array of detectors.
The probablity for a tau neutrino with energy $E_{\nu}$ which is coming from a
 direction of nadir angle $\theta$ to survive after travelling a distance $X$
 is

\beq
P_{surv,\nu_{\tau}}=exp\left[{-\int_0^X \frac{dX^{\prime}}{L_{CC}^{\nu}(E_{\nu_{\tau}},\theta,X^{\prime})}}\right].
\eeq

In the above expression 
$$L_{CC}^{\nu}(E_{\nu_{\tau}},\theta,X)=[\sigma_{CC}^{\nu}(E_{\nu_{\tau}})\rho[r(\theta,X)]N_A]^{-1}.$$ is the charge current interaction length. $\sigma_{CC}^{\nu}(E_{\nu_{\tau}})$ is the neutrino nucleon charge current interaction cross section. Here the interactions are $$\nu_{\tau}(\bar{\nu_{\tau}})+N(nucleon) \rightarrow \tau^{-}(\tau^{+}) +anything.$$ 
The tau neutrino interacts with a nucleon in the matter of Earth. $\rho(r)$ is
 the density of Earth at a distance of $r$ from the center of the Earth,
 and $N_A=6.022\times10^{23}$. The distance from the center of Earth is given
 by $$r(\theta,X)= R_{e}^2+X^2-2R_{e}X\cos{\theta}.$$
 $R_{e}$ the radius of the Earth is taken to be $\approx 6371 km$.
We only consider tau neutrinos with energy above $10^{8} GeV$. For
 $E_{\nu_{\tau}}\gsim 10^{8} GeV$ the charge current cross sections for
 $\nu_{\tau}$ and $\bar{\nu_{\tau}}$ are virtually identical. 

\para

The probablity for a tau lepton production from a tau neutrino in a distance 
between $X$ and $X+dX$ is $dX/L_{CC}^{\nu}(E_{\nu_{\tau}},\theta,X)$. We are 
interested only in tau leptons which travel nearly horizontally so that they 
can decay in the atmosphere and produce visible signals in ground array 
detectors and therefore in our case the generation of tau leptons from tau 
neutrinos occur near Earth's surface where Earth's density is
 $\rho_{s}=2.65 gm/cm^{3}$. The differential probality of conversion for tau
 neutrinos to tau leptons can be expressed as
\beq
dP_{conv}=\frac{dX}{L_{CCs}^{\nu}(E_{\nu_{\tau}})}.
\eeq

The expression for charge current interaction length gets simplified in this
 case because the tau neutrino beam is passing the surface of Earth almost
 horizontally. 
$L_{CCs}^{\nu}(E_{\nu_{\tau}})=[\sigma_{CC}^{\nu}(E_{\nu_{\tau}})\rho_{s}N_{A}]^{-1}$.
 
In our calculation we have used the result from \cite{gandhi} that
the tau leptons take $80\%$ of the tau neutrino energy.
 A charged tau lepton loses energy as it moves through the Earth. 
The energy loss rate can be parametrised as
\beq
dE_{\tau}/dX=-(\alpha_{\tau}+\beta_{\tau}E_{\tau})\rho[r(\theta,X)] 
\eeq
The above equation parametrises lepton energy loss through bremsstrahlung, 
pair production, and photonuclear interactions under the assumption of uniform
 energy loss. Here
 $\beta_{\tau}\approx 0.8\times 10^{-6} cm^{2}/gm$ \cite{lipari} and the 
effect of $\alpha_{\tau}$ at these neutrino energies is negligible.
The probablity of survival for a charged lepton which is losing energy at the
 rate described above is
\beq
dP_{surv,\tau}/dX=-P_{surv,\tau}/(c t_{\tau}E_{\tau}/m_{\tau}).
\eeq
In the above expression $t_{\tau}$ and $m_{\tau}$ are the lifetime and mass of 
tau lepton respectively, $c$ is the speed of light.
Solving eqn.(14) and (15) one gets the expression of survival probablity of a
 tau lepton assuming a constant density $\rho_{s}$ of the surface of Earth.  
\beq
P_{surv,\tau}=exp\left[\frac{m_{\tau}}{c t_{\tau} \beta_{\tau} \rho_s}\left(\frac{1}{0.8E_{\nu_{\tau}}}-\frac{1}{E_{\tau}}\right)\right]
\eeq
In the above expression, we have used the values of lifetime of a tau lepton 
$t_{\tau}=2.96 \times 10^{-13} sec$ and mass of a tau lepton $m_{\tau}=1.777 GeV$.
A tau lepton, produced from a tau neutrino of energy $E_{\nu_{\tau}}$ at a
 distance of $X^{\prime}$ after the tau neutrino enters the Earth, loses 
energy and then exits the surface of Earth with an energy $E_{\tau}$. The
 condition of consistency of the exit energy of a tau lepton with its original
 energy and location gives

\beq
P_{cond}=\delta(E_{\tau}-0.8E_{\nu_{\tau}}e^{-\beta_{\tau}\rho_{s}(2R_{e}\cos{\theta}-X^{\prime})}).
\eeq

In writing the above expression the Earth's surface has been assumed to have a
 constant density $\rho_{s}$. $\theta$ is the nadir angle at which the neutrino
 is coming to Earth.
The combined effect of all the probablities described in equations
 (12),(13),(16) and (17) can be written as

\beq
K(E_{\nu_{\tau}},\theta;E_{\tau})=\int_{0}^{2R_{e}\cos{\theta}}{P_{surv,\nu_{\tau}} P_{conv} P_{surv,\tau} P_{cond}} dX.
\eeq

The range of a tau lepton in Earth is far less than the typical interaction length of a neutrino and therefore the kernel descibed above is dominated by the contribution from $X\approx 2R_{e}\cos{\theta}$.
In equation (12) $X$ is replaced by $2R_{e}\cos{\theta}$.
The expression in equation(17) can be simplified after doing the integration in $X$ using $\int {dX \delta[h(X)]}=\vert{dh/dX}\vert ^{-1}_{h=0}$. 

\beq
K(E_{\nu_{\tau}},\theta;E_{\tau})\approx \frac{1}{L_{CCs}^{\nu}(E_{\nu_{\tau}})}
e^{-\int_{0}^{2R_{e}\cos{\theta}}dX^{\prime}/[L_{CC}^{\nu}
(E_{\nu_{\tau}},\theta,X^{\prime})]}\times exp\left[\frac{m_{\tau}}{ct_{\tau}\beta_{\tau}\rho_{s}}\left(\frac{1}{0.8E_{\nu_{\tau}}}-\frac{1}{E_{\tau}}\right)\right] \frac{1}{E_{\tau}\beta_{\tau}\rho_{s}}.  
\eeq
    
We fold this kernel with the tau neutrino spectrum on Earth and the aperture of the detector to calculate the number of tau lepton events or signals in the detector.

\beq
{M}_{\tau}=\int dE_{\nu_{\tau}} dE_{\tau} \frac{d{M}_{\nu_{\tau}}}{dE_{\nu_{\tau}}}K(E_{\nu_{\tau}},\theta,E_{\tau}) (A\Omega)_{eff}
\eeq

\para
We integrate over tau neutrino energies from $10^{8}GeV$ to $10^{11}GeV$ 
to calculate the number of visible tau events in the atmosphere.The signal 
from an electromagnetic shower competes with the background noise from the 
night sky. It is possible to calculate the energy required for an 
electromagnetic shower to be detectable by considering the signal to noise
 ratio in individual photomultiplier tubes. The effective apertures for
 earth-skimming tau leptons for HiRes and Telescope Array are given
 in \cite{feng}. 
These apertures are calculated for diffuse sources of tau leptons (these tau
 leptons are coming from all directions). The apertures are expressed in units 
of $km^2 sr$.

However, in our work we consider individual GRBs as the source of neutrinos and
 therefore the effective apertures given in \cite{feng} have to be modified to
 use them for a point source. The point source would be observable only if it
 is in the field of view of the detector and not otherwise. 
The azimuthal angular coverage for HiRes is nearly $360^{\circ}$ \cite{hires} 
and for Telescope Array also it is $360^{\circ}$ \cite{tel1}, which means that
 tau leptons coming from directions covering $360^{\circ}$ will be detectable by 
HiRes and Telescope Array. But only those tau leptons which travel within 0.3 
rad from the surface of the Earth will produce visible signals in the 
atmosphere.  So tau signals which are coming within 
$360^{\circ}\times (0.3/{\pi}) 180^{\circ}$ to HiRes or Telescope Array will
 be detectable by them.

Dividing the effective apertures given in \cite{feng} by the solid angle
 covered by these detectors for detecting high energy $\tau$ leptons passing
 close to the surface of the Earth, we find the effective apertures for
 tau leptons produced from neutrinos emitted by $\lq{point}\rq$ sources. 

\section{Results and Discussions}
We have calculated the expected number of tau neutrino events from  individual
 GRBs for detectors like HiRes and Telescope Array.
These tau neutrinos are coming to the detectors almost horizontally and
 producing visible signals in the atmosphere. In the present work we have 
assumed that the burst is occuring at a nadir angle of $\theta=89^{\circ}$
 and we have considered the neutrinos which have energies from
 $10^{8}GeV$ to $10^{11}GeV$ to calculate the total number of expected tau
 events from the GRB. 

\para
 
Each ground array of detectors has a duty cycle. Each GRB is usually of a
 few seconds duration. 
Only those bursts whose occurence near the horizon coincides with
the requirement of clear moonless nights for fluorescence detection will be
 detectable by the detectors.

\para

In Figure 1 we plot the expected number of tau neutrinos from a GRB in a generic
 detector of $1 km^{2}$ area. One should note that this figure does not show 
results for HiRes or Telescope Array. While plotting the number of tau
 neutrinos against the Lorentz factor $\Gamma$ of the GRB, we have varied the
 threshold energy of the detector. We have assumed that $\alpha=0.01,
\beta=1.2,z=0.5,E_{\nu,HE}=10^{51}erg$. The GRB is assumed to 
occur at nadir angle of $89^{\circ}$.
The tau neutrino spectrum is described by three functional forms in three
 ranges of energies in eq.2 of our text. If the threshold energy of the
 detector is above the neutrino break energy $E_{\nu}^{b}$ then the first
 part of the spectrum does not contribute to the total number of tau neutrinos. 
The break energies of the neutrino spectrum are proportional to
 ${\Gamma}_{2.5}^4$ and hence a slight change in $\Gamma$ can produce
 significant changes in the break energies as well as in the number of tau
 neutrinos. This is the reason why there is an extreme jump in the number of
 tau signals for $E_{th}=10^{8} GeV$ when it is displayed as a function of
 $\Gamma$.

\para

In Figure 2 we have plotted the expected number visible tau signals
 against Lorentz factor of a GRB for HiRes detector. The burst is assumed to be
 occuring at a nadir angle $\theta=89^{\circ}$. We assume that the burst is
 occuring at $z=0.5$ and the total energy emitted by the GRB in
 muon neutrinos and muon antineutrinos is $E_{\nu,HE}=10^{51}erg$. The 
luminosity of the burst is $L_{\gamma}=10^{51}erg/sec$ and the spectral indices
 of the photon spectrum are $\alpha=0.01$ and $\beta=1.2$. The values of the
 equipartition parameters are assumed to be $\epsilon_{e}=0.45$ and 
$\epsilon_{B}=0.1$. 
The break energies of the neutrino spectrum $E_{\nu}^{b}$ and
 $E_{\nu,\bar{\nu}}^{s}$ are proportional to ${\Gamma^{4}_{2.5}}$. In the
 figure the changes in the shape of the visible tau spectrum are due to the 
changes in the break energies of the spectrum of neutrinos as we vary $\Gamma$. 
Also, the break energies vary linearly with the variability time $t_{v}$. 
As we increase $\Gamma$ the number of signals initially increase and after
 reaching a peak value it falls down. The effect of the  variation in $t_{v}$
on the number of visible tau lepton signals is also shown.  
  
\para

The redshift of a GRB has also an important role in determining the number of
 signals from that GRB. Figure 3 shows how the number of visible signals from 
a GRB falls down drastically as the distance of the GRB from us increases. We 
have assumed $\alpha=-0.3$ and considered three cases $\beta=1.2$, 
$1.5$ and $2$.
 The Lorentz factor $\Gamma$ of the GRB is assumed to be $300$. The total 
energy emitted in muon neutrinos and antineutrinos is $E_{\nu,HE}=10^{51}erg$
 and $L_{\gamma}=10^{51}erg/sec$ as in Figure 1.

\para

The next figure shows the role of $\alpha$ in determining the number of visible
 tau signals in HiRes. Comparing Figure 3 and 4 we find that the 
variation in number of tau signals is more due to the variation in the value 
of $\alpha$ than due to the variation in the value of $\beta$.
 
\para

In Figure 5 we observe the dependence of the number of tau signals on the 
equipartition parameters $\epsilon_{e}$ and $\epsilon_{B}$.
The equipartition parameteres enter in the expressions of break energies of the
 neutrino spectrum. The variation in the number of visible tau signals is more
 due to the variation in the value of $\epsilon_{e}$ than due to the variation
 in the value of $\epsilon_{B}$ because the neutrino spectrum break energy 
$E_{\nu}^{b}$ is proportional to $\epsilon_{e}^{-3/2}$ and
 $\epsilon_{B}^{-1/2}$. 

\para

In Figure 6 and 7 we have calculated the number of visible tau signals
 expected to be detected by Telescope Array. Here we have assumed 11 and 2
 detectors for Telescope Array and HiRes respectively as in \cite{feng}. In
 Figure 7 as we vary $L_{\gamma}$, the break energies of the neutrino spectrum
 change and it is reflected in the number of visible tau lepton signals from 
 the GRB.

\para
The number of tau lepton signals increases linearly with an increase in
 $E_{\nu,HE}$.
\para

Since most of the GRBs occur at high redshifts $(z>0.5)$, from Figure.3
it can be seen that there is almost no chance of observing visible tau signals
 near the horizon with HiRes detector. However, there is a possibility of
 detecting these signals from GRBs with the Telescope Array detector. If the
 GRB has a Lorentz factor about $300$, it emits energy in neutrinos of the
 order of $10^{51}erg$, we would expect to detect tau signals from it with
Telescope Array even if it occurs at a redshift of about 0.5. 
 
But the important point is that the geometrical configuration to produce a
 visible tau signal must be met. The neutrinos should come to the detector
 from the source almost in a  horizontal direction and the leptons would decay 
in the atmosphere to produce visible signals.

We can study neutrino oscillations in astrophysical neutrinos, production of
 ultra high energy cosmic rays by GRBs and verify models of high energy 
neutrino production inside a GRB, if we can detect visible tau signals from a
 GRB by Telescope Array or other similar detectors in future.  
\section{Conclusions}
Our calculations show that HiRes is not capable of detecing visible tau signals at
 a nadir angle of $89^{\circ}$ from a GRB of Lorentz factor $300$ even at a 
redshift of 0.2, which is emitting energy in neutrinos of the order of 
$10^{51}erg$.
 If the burst is occuring at redshift $z=0.5$, emitting energy $10^{51} erg$
 in neutrino emissions and its $\Gamma$ is 300, it should be able to produce an 
observable tau lepton signal in the Telescope Array detector.
The number of tau signals depends on the Lorentz factor of the GRB $\Gamma$, 
the redshift $z$ of the GRB, energy emitted in neutrinos $E_{\nu,HE}$,
 variability time $t_{v}$, photon luminosity $L_{\gamma}$, equipartition
 parameters $\epsilon_{e}$, $\epsilon_{B}$ and also on the photon spectral 
indices. There is an opportunity of understanding neutrino oscillations in 
astrophysical neutrinos, production of ultra high energy cosmic rays by GRBs 
and models of high energy neutrino production inside a GRB by detecting
 visible tau signals from it.
\section{Acknowledgment}
The author wishes to thank J. L. Feng for helpful communication. A part of
 this work has been done during the author's visit to ICTP, Italy. The author 
is thankful to ICTP for hospitality and providing facilities for her research
 work.

\newpage

\begin{figure}
\begin{center}
\centerline{
\epsfxsize=25. cm\epsfysize=25. cm
\epsfbox{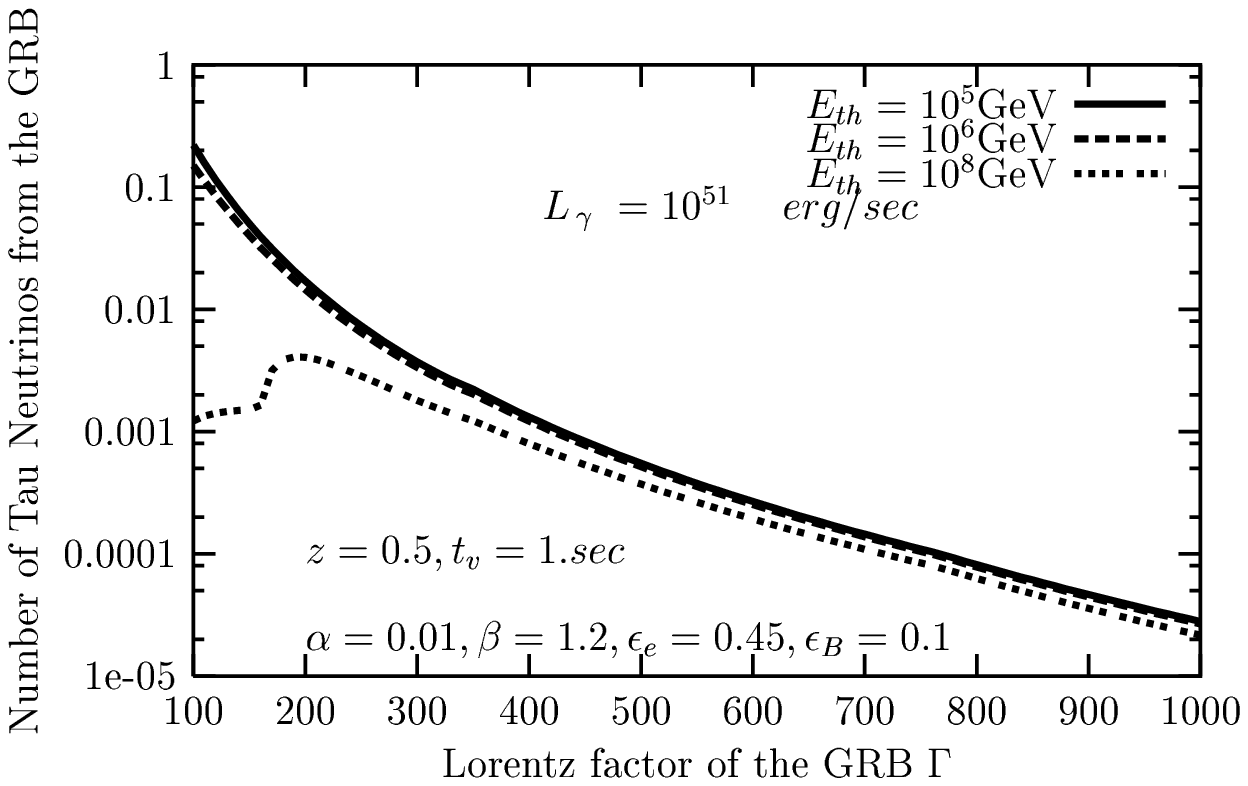} 
}
\end{center}
\vskip - 14 cm
\caption{The expected number of tau neutrinos from a GRB in a detector of
 $Km^2$ area has been plotted against the Lorentz factor of that GRB. We have
 varied the threshold energy $E_{th}$ of the detector to show how the number
 of tau neutrinos in the detector changes due to the change in the threshold 
energy of the detector. The GRB is observed at a nadir angle of $89^{\circ}$
 and at a redshift $z=0.5$. We assume $\epsilon_{e}=0.45$, $\epsilon_{B}=0.1$
 and $E_{\nu,HE}=10^{51}erg$.}
\end{figure}

\newpage
\begin{figure}
\begin{center}
\centerline{
\epsfxsize=25. cm\epsfysize=25.cm
\epsfbox{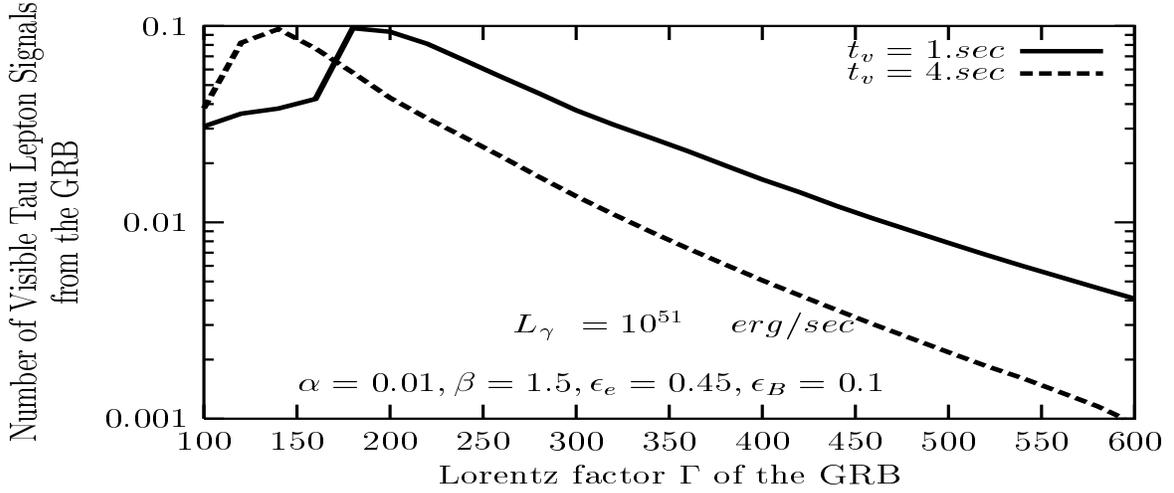} 
}
\end{center}
\vskip -14 cm
\caption{The expected number of visible tau lepton events from a GRB has been
 plotted against its Lorentz factor $\Gamma$ for HiRes detector. The burst is
 assumed to occur at $\theta=89^{\circ}$. The variability time of the GRB has 
been varied. We assume $z=0.5$, $E_{\nu,HE}=10^{51}erg$, 
$L_{\gamma}=10^{51}erg/sec$, $\epsilon_{e}=0.45$, $\epsilon_{B}=0.1$,
 $\alpha=0.01$ and $\beta=1.5$.}
\end{figure}

\newpage
\begin{figure}
\begin{center}
\centerline{
\epsfxsize=25. cm\epsfysize=25. cm
\epsfbox{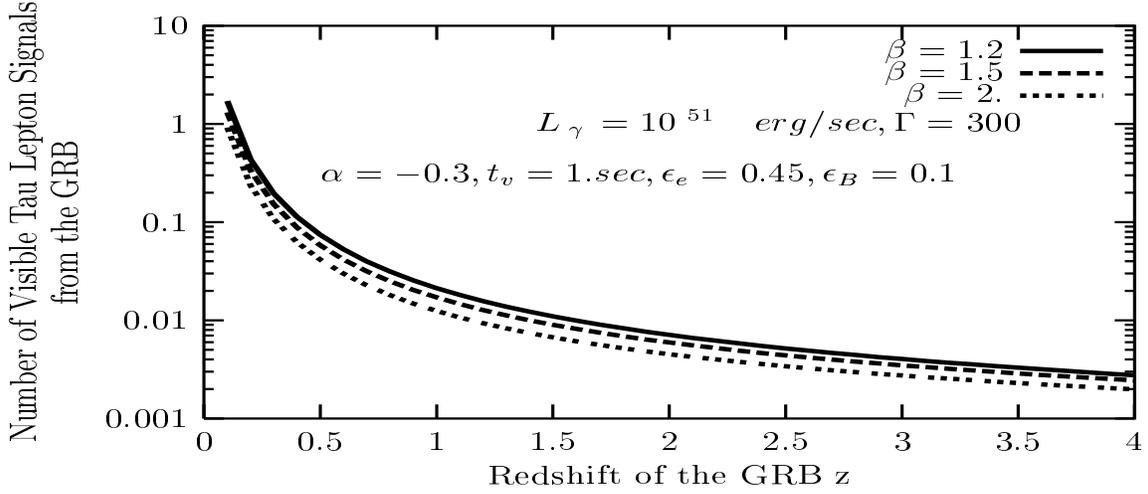} 
}
\end{center}
\vskip -14 cm
\caption{This figure shows how the expected number of visible tau lepton
 events falls of as we go to higher redshifts. The burst is assumed to be
 observed at a nadir angle of $\theta=89^{\circ}$ with HiRes detector. Lorentz 
factor of the GRB is assumed to be $\Gamma=300.$ and total energy emitted by 
the GRB is $E_{\nu,HE}=10^{51}erg$. We consider three values of $\beta$ in our
 plot $\beta=1.2,1.5,2$. We also assume $\alpha=-0.3$, $\epsilon_{e}=0.45$, 
$\epsilon_{B}=0.1$, $t_{v}=1.sec$. The luminosity of the burst is assumed to 
be $L_{\gamma}=10^{51}erg/sec$.}
\end{figure}

\newpage
\begin{figure}
\begin{center}
\centerline{
\epsfxsize=25. cm\epsfysize=25. cm
\epsfbox{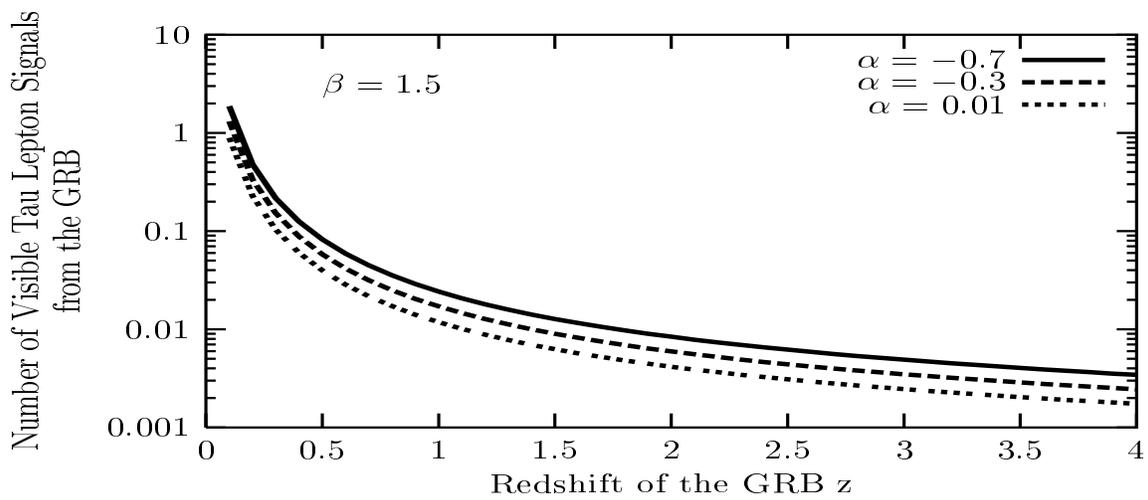} 
}
\end{center}
\vskip -14 cm
\caption{This figure is same as figure.3. The only difference is that in this
 case we assume $\beta=1.5$ and we consider three cases $\alpha=-0.7$,$-0.3$ 
and  $0.01$.}
\end{figure}

\newpage
\begin{figure}
\begin{center}
\centerline{
\epsfxsize=25. cm\epsfysize=25.0cm
\epsfbox{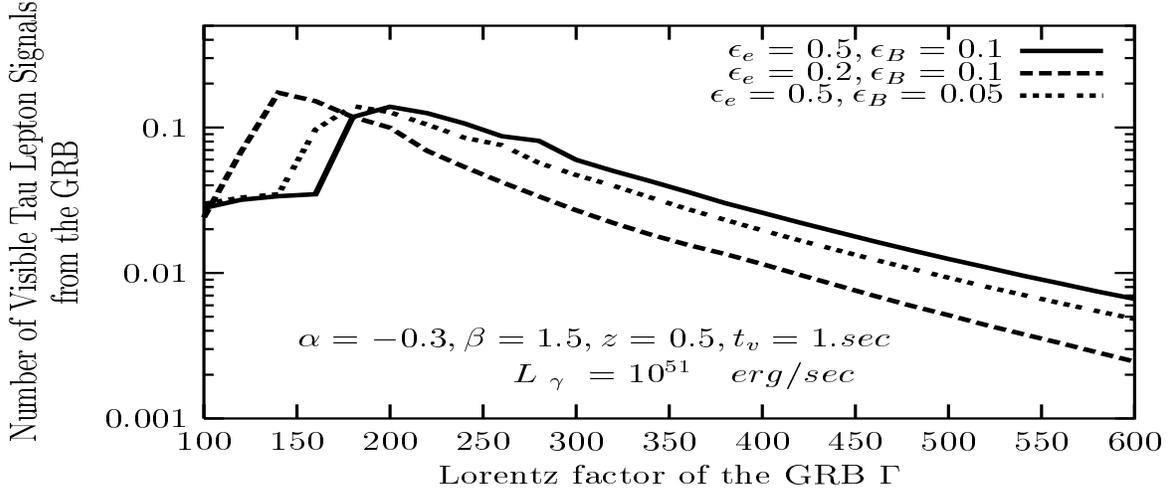} 
}
\end{center}
\vskip -14 cm
\caption{The expected number of visible tau signals has been plotted against
 Lorentz factor $\Gamma$ of the GRB. The GRB is assumed to be observed at 
nadir angle $\theta=89^{\circ}$ with HiRes. We assume redshift of the GRB is
 $z=0.5$, the total energy emitted initially in muon neutrinos is 
$E_{\nu,HE}=10^{51}erg$ and $\alpha=-0.3$, $\beta=1.5$. The variability time 
is assumed to be $t_{v}=1.sec$. }   
\end{figure}

\newpage

\begin{figure}
\begin{center}
\centerline{
\epsfxsize=25. cm\epsfysize=25. cm
\epsfbox{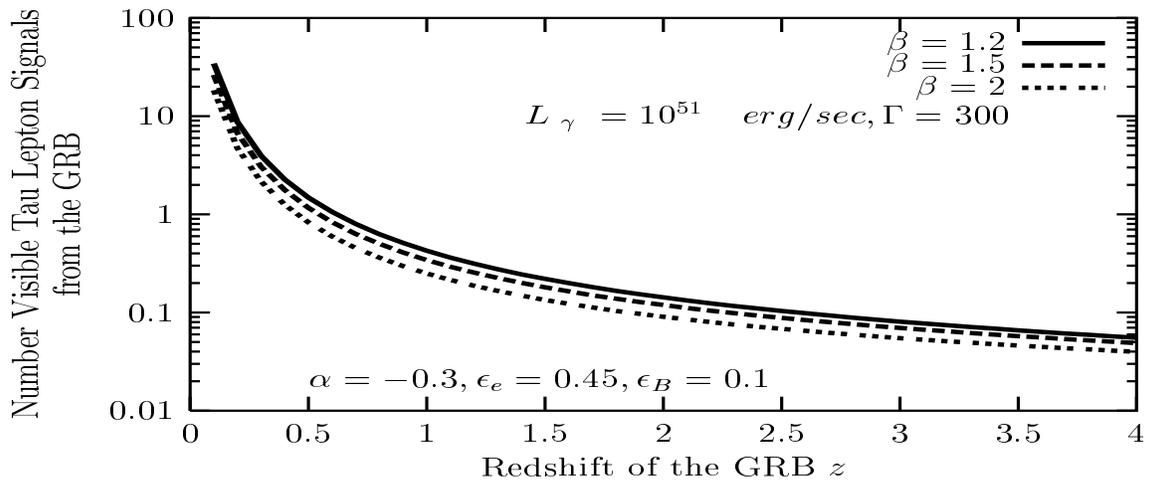}
}
\end{center}
\vskip -14 cm
\caption{The expected number of visible tau lepton events in the atmosphere 
has been plotted for Telescope Array against redshift of the GRB. The burst is
 assumed to be observed near the horizon at nadir angle $\theta=89^{\circ}$. We
 assume $\alpha=-0.3$, $\Gamma=300.$, $E_{\nu,HE}=10^{51}erg$, $ t_{v}=1.sec$, 
$\epsilon_{e}=0.45$, $\epsilon_{B}=0.1$ and $L_{\gamma}=10^{51}erg/sec$}   
\end{figure}

\newpage
\begin{figure}
\begin{center}
\centerline{
\epsfxsize=25.cm\epsfysize=25.cm
\epsfbox{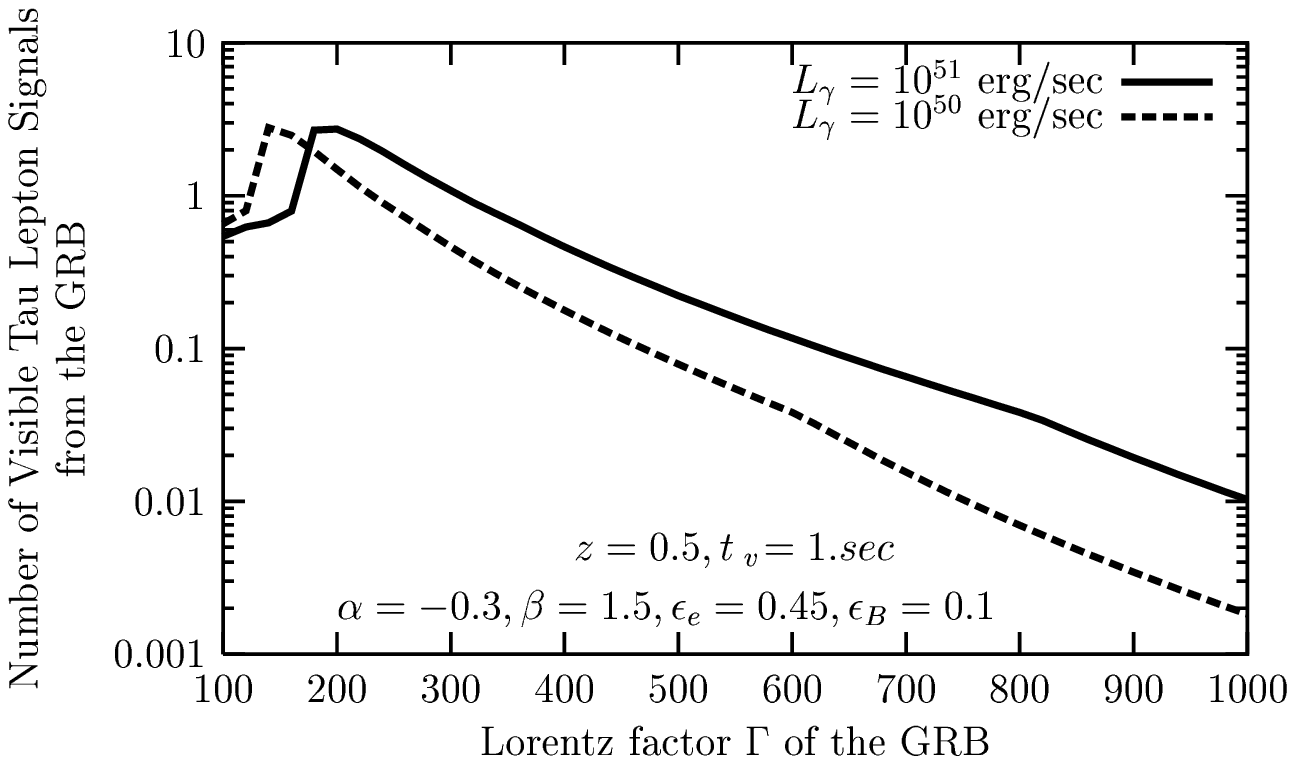}
}
\end{center}
\vskip -14 cm
\caption{This figure shows the dependence of expected number of visible tau
 signals on the Lorentz factor of the GRB $\Gamma$. The burst is assumed to be
 observed at nadir angle $\theta=89^{\circ}$ as earlier with Telescope Array.
 We consider two cases $L_{\gamma}=10^{50}erg/sec$ and 
$L_{\gamma}=10^{51}erg/sec$. The other parameters are $z=0.5$, 
$E_{\nu,HE}=10^{51}erg$, $\alpha=-0.3$, $\beta=1.5$, $t_{v}=1.sec$, 
$\epsilon_{e}=0.45$ and $\epsilon_{B}=0.1$.} 
\end{figure}

\newpage
\begin{thebibliography}{99}
\bibitem{nayan1} N. Gupta, Phys. Lett. B {\bf 541}, 16 (2002). 
\bibitem{fargion} D. Fargion, ApJ {\bf 570}, 909 (2002);
 and astro-ph/0101565, Vulcano DARK 2000 conference.
\bibitem{selvon} A. Letessier-Selvon, astro-ph/0009444. 
\bibitem{bertou} X. Bertou, P. Billoir, O. Deligny, C. Lachaud and A. Letessier-Selvon, Astropart. Phys. {\bf 17}, 183 (2002).
\bibitem{feng} J. L. Feng, P. Fisher, F. Wilczek and T. M. Yu, Phys. Rev. Lett. {\bf 88}, 161102 (2002). 
\bibitem{sasaki} M. Sasaki, Y. Asaoka and M. Jobashi, ICRR-Report-484-2002-2.
\bibitem{eichler} D. Eichler et al., Nature {\bf 340}, 126 (1989).  
\bibitem{narayan} R. Narayan, B. Paczy\'{n}ski and T. Piran, ApJ {\bf 395}, L83 (1992).
\bibitem{fryer1} C. Fryer and S. E. Woosley, ApJ {\bf 502}, L9 (1998). 
\bibitem{fryer2} C. Fryer, S. E. Woosley and D. H. Hartmann, ApJ {\bf 526}, 152 (1999).
\bibitem{woosley} S. E. Woosley, ApJ {\bf 405}, 273 (1993).
\bibitem{pacy} B. Paczy\'{n}ski, ApJ {\bf 494}, 45 (1998).
\bibitem{vietri} M. Vietri and L. Stella, ApJ {\bf 507}, L45 (1998).
\bibitem{rees} M. J. Rees and P. M\'esz\'aros, MNRAS, {\bf 258}, 41P (1992).
\bibitem{granot}J. Granot, T. Piran and Re'em Sari, ApJL {\bf 534}, L163 (2000).
\bibitem{waxman} E.Waxman, J. N. Bahcall, Phys. Rev. Lett. {\bf 78}, 2292 (1997). E. Waxman, 2003, astro-ph/0303517, {\it ``Supernovae and Gamma Ray Bursters'' ed. K. W. Weiler, Lecture Notes in Physics, Springer Verlag} (in press).
\bibitem{vietri1} M. Vietri, D. D. Marco, D. Guetta, astro-ph/0302144.
\bibitem{hires} T. Abu-Zayyad et al., 2002, astro-ph/0208243.
\bibitem{agasa} M. Takeda et al., Phys. Rev. Lett. {\bf 81}, 1163 (1998). 
\bibitem{auger} The Pierre Auger Project Design Report, Fermilab (1995). www.auger.org.  
\bibitem{guetta1} D. Guetta, M. Spada and E. Waxman, ApJ {\bf 559}, 101 (2001).
\bibitem{waxman1} E. Waxman, ICTP Summer School on Astroparticle Physics and Cosmology, Italy, June 2000 and VI Gleb Wataghin School on High Energy Phenomenology (UNICAMP, Campinas Brazil, July 2000),
\bibitem{guetta} D. Guetta, D. Hooper, J. Alvarez-Mu\~niz, F. Halzen and E. Reuveni  astro-ph/0302524.
\bibitem{wmap} C. L. Bennett et al., 2003, astro-ph/0302207.
\bibitem{frail} D. A. Frail et al., Astrophys. J. Lett. {\bf 562}, L55 (2001).
\bibitem{super}Super-Kamiokande Collaboration, Y. Fukuda et al. Phys. Rev. Lett {\bf 81}, 1562 (1998).
\bibitem{gandhi}R. Gandhi, C. Quigg, M.H. Reno, and I. Sarcevic, Astropart. Phys. {\bf 5}, 81 (1996). 
\bibitem{lipari}P. Lipari and T. Stanev, Phys. Rev. D {\bf 44}, 3543 (1991).
\bibitem{tel1} M. Sasaki et al., OG 4.5.15 $26^{th}$ ICRC, {\bf 5}, 397 (1999).
\end{thebibliography}
\end{document}